\documentclass[reprint,eqsecnum,floats,aps,amsmath,amssymb,nofootinbib,prd,onecolumn, showpacs]{revtex4-1}

\usepackage{graphicx,physics}
\usepackage{amsmath,amssymb,mathtools,mathrsfs}
\usepackage{hyperref}
\usepackage{graphicx}
\usepackage{subfigure}
\usepackage{arydshln}
\usepackage{xcolor,bm}
\usepackage{braket}
\usepackage{tensor}
\usepackage{enumitem,array,textcomp,soul}

\begin{document}
	
\title{Primordial perturbations in the Dapor-Liegener model of hybrid loop quantum cosmology}
\author{Laura Castell\'o Gomar}
\email{laura.castello@iem.cfmac.csic.es}
\affiliation{Instituto de Estructura de la Materia, IEM-CSIC, Serrano 121, 28006 Madrid, Spain}
\author{Alejandro Garc\'ia-Quismondo}
\email{alejandro.garcia@iem.cfmac.csic.es}
\affiliation{Instituto de Estructura de la Materia, IEM-CSIC, Serrano 121, 28006 Madrid, Spain}
\author{Guillermo  A. Mena Marug\'an}
\email{mena@iem.cfmac.csic.es}
\affiliation{Instituto de Estructura de la Materia, IEM-CSIC, Serrano 121, 28006 Madrid, Spain}
\begin{abstract}
In this work, we extend the formalism of hybrid loop quantum cosmology for primordial perturbations around a flat, homogeneous, and isotropic universe to the new treatment of Friedmann-Lema\^itre-Robertson-Walker geometries proposed recently by Dapor and Liegener, based on an alternative regularization of the Hamiltonian constraint. In fact, our discussion is applicable also to other possible regularization schemes for loop quantum cosmology, although we specialize our analysis to the Dapor-Liegener proposal and construct explicitly all involved quantum operators for that case.
\end{abstract}
	
\maketitle
	
\section{Introduction}
	
The theory of primordial perturbations \cite{perturbations1, perturbations2,HalliwellHawking,ShiraiWada} in a spatially homogeneous and isotropic Friedmann-Lema\^{i}tre-Robertson-Walker (FLRW) model is the cornerstone of modern cosmology, since it allows us to connect theories of the very early Universe with the most precise data obtained experimentally. Combined with the inflationary paradigm \cite{inflationmodels1,inflationmodels2,inflationmodels3,inflationmodels4}, this satisfactorily describes the evolution of the Universe from its primeval stages and explains the formation of structures observed nowadays at large scales \cite{inflation1}. The basic idea is that primordial inhomogeneities around a smooth background emerged from quantum vacuum fluctuations, and then provided the seeds of cosmological structure formation under the action of gravitational instability \cite{inflation2}. Therefore, one expects that both quantum mechanics and general relativity will be ultimately required in order to fully understand the generation and evolution of fluctuations in an accurate cosmological description. 
	
The standard model is based on the conceptual framework of quantum field theory (QFT) in a classical and fixed curved spacetime, where perturbations are viewed as quantum test fields propagating on a given geometry. Present observations of the cosmic microwave background (CMB) broadly lead to a solid confirmation of the predictions of this standard cosmology. Even so, there appear to exist some puzzling discrepancies between the theoretical results and the observations for large angular scales (around low multipoles, close to number 30 and below) \cite{data1,data2,data3}. At these scales, however, all measurements are affected by the errors caused by cosmic variance. In the wait for new relevant advances in this front, both the polarization signal of the CMB and the search for signals of the gravitational wave background emitted during the inflationary epoch may provide interesting new frontiers in observational cosmology, as they might offer key information about the early Universe and open a window to new physics.  
	
From a fundamental point of view, the hope is that these data may encode information about phenomena caused by the quantum nature of spacetime geometry itself, as quantum gravity effects should be relevant in the extreme conditions experienced in the early stages of the cosmos. Since a definitive quantum theory of gravity has not yet been established satisfactorily, in the past ten years an original approach has been proposed to deal with the quantization of inhomogeneous gravitational models. The so-called \emph{hybrid quantum cosmology} formalism \cite{hybridGIper} combines two different types of quantum representations: one of them is based on a genuine quantum theory of geometry, used for the description of (at least) the homogeneous sector of the cosmological system, and the other consists in a more conventional Fock quantum description, employed for the inhomogeneities (typically identified with perturbations) of both the  geometry and the matter content. This hybrid approach rests basically on the assumption that there exists a regime of the quantum dynamics, between a fully quantum gravity regime and the scheme of QFT in fixed curved spacetimes, where the most relevant quantum effects of the geometry are those affecting the zero modes of the homogeneous sector. An important point for the application of this formalism to cosmological perturbations is the mathematically consistent truncation of the total action at quadratic order in the perturbations. Indeed, this truncation of the action allows one to maintain a symplectic structure for the whole system, formed by the homogeneous degrees of freedom and the perturbations, while keeping the restrictions imposed on this system by the gravitational constraints. In general, in the implementation of this hybrid formalism, one assumes that the quantization satisfies the canonical commutation relations inasmuch as the operators for the homogeneous (and isotropic) geometry commute with those representing the homogeneous sector of the matter fields and, in turn, all of these commute with the elementary operators corresponding to the variables that describe the inhomogeneities. 

The hybrid approach was first put forward some ten years ago to deal with cosmological universes with compact sections that sustain gravitational waves. More specifically, the approach was introduced to attain a quantum formulation of the so-called Gowdy cosmologies \cite{gowdy1,gowdy2} for the model with three-toroidal spatial sections and linearly polarized waves \cite{hybridgowdy1,hybridgowdy2,hybridgowdy3,hybridgowdy4}. The strategy was soon extended to the analysis of the more realistic system of an FLRW universe with cosmological perturbations, discussing both scalar \cite{hybridperturbations1,hybridperturbations2,hybridperturbations3} and tensor perturbations \cite{tensormodes}. In principle, these original works were specialized to the case where the homogeneous sector of the gravitational system is quantized according to the rules of loop quantum cosmology (LQC) \cite{AS,LQC}. LQC is the particularization to cosmological reductions of general relativity of the strategies of loop quantum gravity (LQG) \cite{ALQG,Thiem}, a nonperturbative formalism for the quantization of the Einsteinian theory.  However, the hybrid approach can be generalized and adapted to other candidates for the quantum description of the homogeneous sector of cosmological spacetimes, although this fact may not have been sufficiently emphasized in the literature. Actually, the discussion carried out in Ref. \cite{hybridGIper} for the gauge invariant treatment of the cosmological perturbations in the hybrid approach was already presented in a way that is suitable for this generalization.

This versatility of the hybrid strategy is especially useful in the situation that is found at present in LQC, where the community is discussing possible alternatives for the regularization of the geometric operators that appear in the analysis of homogeneous cosmologies. In practice, these alternatives, originated from regularization ambiguities \cite{TT,ALM,AAL} in the symmetry reduction to homogeneity, result in different quantization prescriptions for the homogeneous sector of the geometry. In this sense, they can be regarded as different representations of the operators that describe the effect of the zero modes of the geometry. The discussed property of the hybrid approach (namely that it can be accommodated to distinct representations of the zero modes) then allows us to study the quantum inclusion of inhomogeneities while permitting the consideration of any of the mentioned alternatives in LQC. Apart from the standard regularization followed so far in LQC \cite{APS2,MMO}, most of the attention has been focused on a regularization procedure recently put forward by Dapor and Liegener (DL) \cite{DL1}, following pioneering works by Thiemann \cite{TT} and Yang, Ding, and Ma \cite{YDM}. This procedure adopts a different regularization scheme for the Euclidean and Lorentzian parts of the gravitational action, namely the parts without and with quadratic dependence on the extrinsic curvature, respectively \cite{Thiem}, implementing this regularization before proceeding to the symmetry reduction to homogeneity and isotropy. This alternative for the construction of the Hamiltonian of LQC has been considered in several papers in the past two years \cite{Paramc1,Paramc2, DL2,DL3,genericness,Agullo,Haro,DaPorLie}.  The extension to anisotropic universes of the Bianchi type I has also been studied \cite{BianchiDL}. An especially appealing property of the new Hamiltonian constraint obtained in this manner for flat FLRW cosmologies is that the effective solutions typically present two branches with different cosmological behavior: one of them corresponding to an asymptotically de Sitter cosmology, even in the absence of a genuine cosmological constant, and the other one describing an FLRW universe. Therefore, one finds a physical scenario in which a contracting de Sitter regime is followed by a quantum bounce after which there may exist an inflationary era leading to a universe like the one we observe. This suggestive fact, with potential implications for the generation of primordial perturbations, together with the convenience of studying well-founded regularizations in LQC other than the standard one (which seems desirable given that we do not yet fully understand the relation between LQG and this formalism of quantum cosmology \cite{Engle,BK,Engle2,Paw,BEHM}), are the main motivations for our interest in investigating the application of the hybrid approach to inhomogeneities in spacetimes that are described quantum mechanically with the DL proposal. 
	
In more detail, in this work we construct a theoretical framework for the treatment of cosmological perturbations adopting the DL proposal for the quantization of the FLRW sector of the geometry. We consider inhomogeneous perturbations on top of a flat FLRW spacetime with a minimally coupled scalar field, focusing on scalar and tensor perturbations. In general, scalar perturbations are physically the most important ones and, conceptually, the most interesting ones, as they couple to energy density fluctuations, and are ultimately responsible for most of the inhomogeneities and anisotropies in the Universe. This also complicates their mathematical description. Additionally, inflation also generates tensor fluctuations in the spatial metric, that can be viewed as gravitational waves. These are not coupled to any other perturbation at the order of our truncations. However, they do induce fluctuations in the CMB which turn out to be a unique signature of the early epochs of the Universe and offer a valuable window on the physics driving in inflation. To deal with the problem of gauge invariance \cite{Bardeen} at the level of the perturbations while maintaining a canonical set of variables for the entire cosmological model, including the homogeneous FLRW universe, we follow the formulation elaborated in Ref. \cite{hybridGIper}. In particular, the physical degrees of freedom of the scalar perturbations will be described by Mukhanov-Sasaki (MS) variables \cite{S1,S2,M1,M2}, that are perturbative gauge invariants and are directly related to the comoving curvature perturbations. For the tensor perturbations, which are directly perturbative gauge invariant variables, we  adopt the description of Ref. \cite{tensormodes}.
Finally, for the quantization of the FLRW part of the geometry according to the DL proposal, we follow the prescription introduced in Ref. \cite{DaPorLie}, that extends to this new alternative regularization the quantization strategy that was detailed in Ref. \cite{MMO} for the standard regularization employed in LQC.
	
The rest of the paper is organized as follows. The basic results of previous works on the gauge invariant description of the perturbed model are summarized in Sec. \ref{review}. At the end of that section, we specify the resulting zero mode of the Hamiltonian constraint (the only constraint that remains on the system after adopting suitable gauge invariant variables for the perturbations), including terms that are quadratic in the inhomogeneous perturbations. In Sec. \ref{quantum}, we proceed to the quantization of our constrained system (formed by the homogeneous background and the perturbations) following the hybrid approach and adhering to the new formalism of LQC proposed by Dapor and Liegener for the homogeneous geometry. In particular, an alternative representation of the momentum of the scale factor (or, equivalently, of the physical volume) is constructed. In Sec. \ref{effective}, we study a class of solutions to the quantum Hamiltonian constraint for which the dependence on the FLRW geometry and on the perturbations can be separated. We then proceed to derive a master constraint for the gauge invariant perturbations and obtain the time-dependent masses that govern the dynamics of scalar and tensor perturbations. To conclude, we present a summary of the main results and some further discussions in Sec. \ref{conclude}.  We set the speed of light and the reduced Planck constant equal to one throughout our discussion.

\section{Gauge invariant perturbations around a flat FLRW cosmology}\label{review}
	
The theory of cosmological perturbations is complicated by the issue of gauge invariance (see e.g. Ref. \cite{Bardeen}). By performing a small amplitude transformation of the spacetime coordinates, one can introduce fictitious fluctuations around a homogeneous and isotropic spacetime. These fluctuations are \emph{gauge artifacts} that carry no physical significance. There are mainly two approaches to deal with these gauge ambiguities: to perform a gauge fixing or to work directly with gauge invariant variables. The formulation in terms of these gauge invariants is especially convenient for the passage to the quantum theory, since they allow us to reach results that are not by-products of a particular choice of gauge and because they already take into account the effects of the fundamental uncertainties about the gauge sector of the perturbations. In this section, we review the construction of a gauge invariant description of the perturbations around flat FLRW cosmologies and explain how this description can be extended to include the degrees of freedom of the FLRW universes while retaining a canonical formalism for the entire system. This construction follows the discussion presented in Refs. \cite{hybridGIper,tensormodes}. We refer the reader to those articles for further details.

\subsection{The model}\label{model}

We consider the inhomogeneous gravitational system resulting from the introduction of perturbations around a flat FLRW cosmology with compact spatial sections homeomorphic to a three-torus. To achieve a nontrivial dynamics, we introduce a matter content given by a minimally coupled scalar field subject to a generic potential. 
In the following, we focus our attention on perturbations of scalar and tensor type (a natural classification in the limit of continuous modes, that can be reached as explained in Ref. \cite{conti}). At the lowest nontrivial perturbative order, vector perturbations are purely a gauge artifact when no matter vector field is present.  
		
It is convenient to recast the perturbed metric functions and scalar field as expansions in a complete set of scalar and tensor harmonics, after which the dynamics of the perturbations is encoded in the Fourier coefficients of these expansions. The zero modes, which are treated exactly \cite{HalliwellHawking,ShiraiWada,hybridperturbations3,KK}, are regarded as a dynamical homogeneous background that can be parametrized using the homogeneous lapse function $N_0$, the logarithmic scale factor $\alpha$, and the zero mode of the scalar field $\varphi$.  At the level of the action, a truncation of the harmonic expansion at the lowest nontrivial order in the Fourier coefficients (i.e., the quadratic one) results in the following Hamiltonian \cite{hybridGIper,tensormodes}:
\begin{equation}\label{Hamiltonian0}
H=N_0\left[H_{|0}+\sum_{\vec{n},\epsilon}H_{|2}^{\vec{n},\epsilon}+\sum_{\vec{n},\epsilon,\tilde{\epsilon}}{}^TH_{|2}^{\vec{n},\epsilon,\tilde{\epsilon}}\right]+\sum_{\vec{n},\epsilon}g_{\vec{n},\epsilon}\tilde{H}_{|1}^{\vec{n},\epsilon}+\sum_{\vec{n},\epsilon}k_{\vec{n},\epsilon}\tilde{H}_{\_1}^{\vec{n},\epsilon},
\end{equation}
where the labels $\vec{n}\in\mathbb{Z}^3-\{0\}$, $\epsilon=\pm$, and $\tilde{\epsilon}=+,\times$ characterize the scalar and tensor modes. Indeed, $\vec{n}$ designates the wave vector of the mode (its first nonvanishing component is typically restricted to be positive to avoid repetition), $\epsilon$ denotes the parity, and $\tilde{\epsilon}$, that is exclusive to tensor modes, labels the polarization \cite{tensormodes}. 

The lapse $N_0$ and the coefficients $g_{\vec{n},\epsilon}$ and $k_{\vec{n},\epsilon}$ in Eq. \eqref{Hamiltonian0} are Lagrange multipliers and their associated equations of motion merely amount to constraints: the zero mode of the Hamiltonian constraint, $H_{|0}+\sum_{\vec{n},\epsilon}H_{|2}^{\vec{n},\epsilon}+\sum_{\vec{n},\epsilon,\tilde{\epsilon}}{}^TH_{|2}^{\vec{n},\epsilon,\tilde{\epsilon}}$, and the linear perturbative constraints, $\tilde{H}_{|1}^{\vec{n},\epsilon}$ and $\tilde{H}_{\_1}^{\vec{n},\epsilon}$. We then realize that the Hamiltonian is composed by a linear combination of constraints, which reflects the invariance under spatial diffeomorphisms and time reparametrizations that is inherited from the full theory of general relativity. On the one hand, the zero mode of the Hamiltonian constraint is composed by two distinct pieces: the Hamiltonian corresponding to the unperturbed flat FLRW background $H_{|0}$, and two terms that collect quadratic contributions from the inhomogeneities, $H_{|2}^{\vec{n},\epsilon}$ and ${}^TH_{|2}^{\vec{n},\epsilon,\tilde{\epsilon}}$. The homogeneous contribution is given by $H_{|0}=e^{-3\alpha}\left[-\pi_{\alpha}^2+\pi_{\varphi}^2+2e^{6\alpha}\bar{W}(\varphi)\right]/2$, where $\pi_\alpha$ and $\pi_\varphi$ are the momenta canonically conjugate to $\alpha$ and $\varphi$, respectively. In addition, $\bar{W}(\varphi)$ is related with the field potential $W(\varphi)$ by $\bar{W}(\varphi)=\sigma^4 W(\varphi/\sigma)$, where we have defined $\sigma^2=4\pi G/(3l_0^3)$ and $l_0$ is the period of the fundamental cycles of the three-tori, isomorphic to the spatial sections. On the other hand, the two remaining constraints $\tilde{H}_{|1}^{\vec{n},\epsilon}$ and $\tilde{H}_{\_1}^{\vec{n},\epsilon}$ are linear in the scalar perturbations. We notice that there is no perturbative  constraint linear in the tensor perturbations, a fact which can ultimately be traced back to the absence of couplings with tensor matter fields.
	
It is important to remark that the system is symplectic at the discussed perturbative order. The canonical variables that coordinatize the homogeneous phase space are $\{w^a\}_{a=1,2}\equiv\{w_q^a; w_p^a\}_{a=1,2}=\{\alpha,\varphi;\pi_\alpha,\pi_\varphi\}$. In the inhomogeneous sector, two sets can be distinguished: one of them describes the dynamics of the scalar perturbations, for which we use the notation $\{X_l^{\vec{n},\epsilon}\}_{l=1,2,3}\equiv\{X_{q_l}^{\vec{n},\epsilon};X_{p_l}^{\vec{n},\epsilon}\}_{l=1,2,3}$, and the other describes the tensor perturbations, $\{d_{\vec{n},\epsilon,\tilde{\epsilon}};\pi_{d_{\vec{n},\epsilon,\tilde{\epsilon}}}\}$. In both cases, they are composed by the dynamical Fourier coefficients and their associated momenta. 

\subsection{Gauge invariant formalism}\label{GI}
	
As commented above, we wish to describe the perturbations in a gauge invariant manner. The variables that describe the tensor modes turn out to be already gauge invariant in the sense of the Bardeen potentials \cite{Bardeen}. Thus, we only need to focus on the scalar sector. 
	
Perturbative gauge invariants are characterized by being invariant under a perturbative diffeomorphism when the background is regarded as fixed, meaning that, in that situation, they Poisson commute with the generators of perturbative diffeomorphisms: the linear perturbative constraints, $\tilde{H}_{|1}^{\vec{n},\epsilon}$ and $\tilde{H}_{\_ 1}^{\vec{n},\epsilon}$. Following Ref. \cite{hybridGIper}, we introduce the so-called Mukhanov-Sasaki (MS) variables $v_{\vec{n},\epsilon}$ \cite{S1,S2,M1,M2}, which are perturbative gauge invariants defined by linear combinations of the configuration variables of the scalar sector $X_{q_l}^{\vec{n},\epsilon}$. Then, we can construct a complete set of compatible, gauge invariant variables for the scalar perturbations provided that we Abelianize the algebra of perturbative constraints. This can be achieved at the considered truncation order by replacing $\tilde{H}_{|1}^{\vec{n},\epsilon}$ with a suitably redefined perturbative constraint $\breve{H}_{|1}^{\vec{n},\epsilon}$ (see Ref. \cite{hybridGIper}). This procedure leads to a Hamiltonian of the same form, once the lapse function is also redefined, $N_0\to \breve{N}_0$, by including in it terms quadratic in the scalar perturbations \cite{hybridGIper}. Appropriate variables canonically conjugate to the ones described above can be found with relative ease \cite{hybridGIper}, completing the change of perturbative variables into a canonical transformation for the inhomogeneous sector: $X_l^{\vec{n},\epsilon}\mapsto V_l^{\vec{n},\epsilon}\equiv\{V_{q_l}^{\vec{n},\epsilon};V_{p_l}^{\vec{n},\epsilon}\}$, where $\{V_{q_l}^{\vec{n},\epsilon}\}_{l=1,2,3}=\{v_{\vec{n},\epsilon},C_{|1}^{\vec{n},\epsilon},C_{\_1}^{\vec{n},\epsilon}\}$ are the new configuration variables and $\{V_{p_l}^{\vec{n},\epsilon}\}_{l=1,2,3}=\{\pi_{v_{\vec{n},\epsilon}},\breve{H}_{|1}^{\vec{n},\epsilon},\tilde{H}_{\_1}^{\vec{n},\epsilon}\}$ are the new momenta. Notice that we have included the perturbative constraints as momentum variables, with conjugate variables $C_{|1}^{\vec{n},\epsilon}$ and $C_{\_1}^{\vec{n},\epsilon}$ that are pure gauge. This facilitates their quantum implementation, as we discuss in Sec. \ref{quantum}.

\subsection{Redefinition of the gauge invariant variables}\label{MS}
	
Before reintroducing the dynamics of the homogeneous background, we want to address an extra freedom that exists in the formalism: the perturbation variables are not uniquely fixed and, indeed, we can perform transformations that leave the canonical structure of the perturbations invariant while changing those variables. In this subsection we discuss a criterion that allows us to eliminate this freedom up to unitary transformations.
	
In Sec. \ref{quantum}, we perform a Fock quantization of the MS and tensor fields. If we carry out a partial reduction (and time deparametrization) of the system, the MS and tensor fields can actually be interpreted as fields propagating in an ultrastatic compact background. There exist a series of works that guarantee the uniqueness, up to unitary equivalence, of the Fock quantization of fields that propagate in such spacetimes and satisfy a dynamical equation of the Klein-Gordon type, with a mass that can be time-dependent \cite{U1,U2,U3,U4,U5,U6}. This result holds when the Fock representation is required to exhibit the two following properties: (i) the vacuum is invariant under the isometries of the spatial sections and (ii) the quantum evolution associated with the Klein-Gordon equation can be implemented unitarily. Therefore, the background symmetries and the unitary implementability of the dynamical (Heisenberg) evolution pick out a single family of Fock representations that are unitarily equivalent. However, for this statement to apply to the case under consideration, the scalar and tensor gauge invariant fields (as well as their associated momenta) need a suitable transformation, that effectively fixes the freedom mentioned above up to unitary equivalence.

Once we perform the rescaling $d_{\vec{n},\epsilon,\tilde{\epsilon}}\to \tilde{d}_{\vec{n},\epsilon,\tilde{\epsilon}}=e^{\alpha}d_{\vec{n},\epsilon,\tilde{\epsilon}}$ (and the inverse scaling $\pi_{d_{\vec{n},\epsilon,\tilde{\epsilon}}}\to\pi_{\tilde{d}_{\vec{n},\epsilon,\tilde{\epsilon}}}$ to preserve the canonical structure), the fields defined by the scalar and tensor modes do in fact satisfy a Klein-Gordon equation of the desired form. Nonetheless, we can still redefine the momentum variable of each gauge invariant mode by adding a term proportional to the corresponding configuration mode, up to a multiplicative function of the homogeneous background. This ambiguity in the definition of $V_{p_1}^{\vec{n},\epsilon}$ and $\pi_{\tilde{d}_{\vec{n},\epsilon,\tilde{\epsilon}}}$ can be removed by requiring that such a function of the homogeneous background be chosen so that the Hamiltonian no longer contains linear terms in the momentum variables. Actually, this is a necessary and sufficient condition for the unitary implementability of the quantum dynamics. Hence, this criterion resolves the ambiguity in the definition of the perturbation variables up to unitary transformations. For the sake of simplicity, in the rest of our discussion we denote the modified momentum variables using the same notation $V_{p_1}^{\vec{n},\epsilon}$ and $\pi_{\tilde{d}_{\vec{n},\epsilon,\tilde{\epsilon}}}$ as above.

\subsection{Canonical transformation in the homogeneous sector}
	
In the previous subsections, we regarded the background variables $\{w^a\}_{a=1,2}$ as fixed, in order to concentrate our attention on the treatment of the inhomogeneous sector. In this way, we determined a canonical transformation in that sector, $\{X_l^{\vec{n},\epsilon},d_{\vec{n},\epsilon,\tilde{\epsilon}},\pi_{d_{\vec{n},\epsilon,\tilde{\epsilon}}}\} \mapsto \{V_l^{\vec{n},\epsilon},\tilde{d}_{\vec{n},\epsilon,\tilde{\epsilon}},\pi_{\tilde{d}_{\vec{n},\epsilon,\tilde{\epsilon}}}\}$. Now, we proceed to extend this transformation to include the background variables, that then also undergo a modification $\{w^a\} \mapsto \{\tilde{w}^a\}\equiv \{\tilde{\alpha},\tilde{\varphi};\pi_{\tilde{\alpha}},\pi_{\tilde{\varphi}}\}$.
	
We want a transformation that leaves invariant the symplectic structure, the information of which is encoded in the Legendre term of the action, that we call $K$. We seek a set of background variables $\{\tilde{w}^a\}_{a=1,2}$ such that $K$ retains its canonical form (up to terms of order higher than quadratic in the perturbations) when expressed in terms of our new variables. As shown in Ref. \cite{hybridGIper} (see also Refs. \cite{pinto1,pinto2}), the result is that the homogeneous background variables receive corrections that are quadratic in the perturbations. 

By reexpressing the zero mode of the Hamiltonian in terms of the variables $\{\tilde{w}^a,V_l^{\vec{n},\epsilon},\tilde{d}_{\vec{n},\epsilon,\tilde{\epsilon}},\pi_{\tilde{d}_{\vec{n},\epsilon,\tilde{\epsilon}}}\}$, new contributions appear in it that are of quadratic perturbative order. It is clear that, since $H_{|2}^{\vec{n},\epsilon}$ and ${}^{T}H_{|2}^{\vec{n},\epsilon,\tilde{\epsilon}}$ are already quadratic, they adopt the same expression at this order of perturbative truncation. Nonetheless, this is not the case for $H_{|0}$: it receives second-order corrections from our change of variables. These corrections effectively add to the terms $H_{|2}^{\vec{n},\epsilon}$ and ${}^{T}H_{|2}^{\vec{n},\epsilon,\tilde{\epsilon}}$. Writing the homogeneous contribution to the constraint as $H_{|0}=e^{-3\tilde{\alpha}}(\pi_{\tilde{\varphi}}^2-\mathcal{H}_0^{(2)})/2$, where $\mathcal{H}_0^{(2)}=\pi_{\tilde{\alpha}}^2-2e^{6\tilde{\alpha}}\bar{W}(\tilde{\varphi})$, the zero mode of the Hamiltonian constraint turns out to be given by $e^{-3\tilde{\alpha}}\tilde{H}$, where \cite{hybridGIper,tensormodes}
\begin{equation}\label{Htilde}
\tilde{H}=\dfrac{1}{2}\left(\pi_{\tilde{\varphi}}^2-\mathcal{H}_{0}^{(2)}-\Theta_e^{S}-\Theta_o^{S} \pi_{\tilde{\varphi}}-\Theta^T\right).
\end{equation}
The $\Theta$-functions introduced in this equation are defined as
\begin{align}
\Theta_o^S&=-\vartheta_o\sum_{\vec{n},\epsilon}(V_{q_1}^{\vec{n},\epsilon})^2,\quad\Theta_e^S=-\sum_{\vec{n},\epsilon}\left[(\vartheta_e\omega_n^2+\vartheta_e^q)(V_{q_1}^{\vec{n},\epsilon})^2+\vartheta_e(V_{p_1}^{\vec{n},\epsilon})^2\right],\\
\Theta^T&=-\sum_{\vec{n},\epsilon,\tilde{\epsilon}}\left[(\vartheta_e \omega_n^2+\vartheta_T^q)(\tilde{d}_{\vec{n},\epsilon,\tilde{\epsilon}})^2+\vartheta_e (\pi_{\tilde{d}_{\vec{n},\epsilon,\tilde{\epsilon}}})^2\right],
\end{align}
with $\omega_n^2=-4\pi^2|\vec{n}|^2/l_0^2$ and
\begin{align}
\vartheta_o&=-12e^{4\tilde{\alpha}}\bar{W}'(\tilde{\varphi})\dfrac{1}{\pi_{\tilde{\alpha}}},\qquad
\vartheta_e=e^{2\tilde{\alpha}},\label{o}\\
\vartheta_e^q&=e^{-2\tilde{\alpha}}\mathcal{H}_0^{(2)}\left(19-18\dfrac{\mathcal{H}_0^{(2)}}{\pi_{\tilde{\alpha}}^2}\right)+e^{4\tilde{\alpha}}[\bar{W}''(\tilde{\varphi})-4\bar{W}(\tilde{\varphi})],\label{eq}\\
\vartheta_T^q&=e^{-2\tilde{\alpha}}\mathcal{H}_0^{(2)}-4e^{4\tilde{\alpha}}\bar{W}(\tilde{\varphi}),\label{Tq}
\end{align}
that do \emph{not} depend on $\pi_{\tilde{\varphi}}$. Here, the prime denotes the derivative with respect to $\tilde{\varphi}$. The quadratic contributions to the zero mode of the Hamiltonian constraint are obviously gauge invariant. The sums of these quadratic terms for the scalar and tensor perturbations are usually called the MS Hamiltonian and the tensor Hamiltonian, respectively. We emphasize as well that there is no linear contribution from $V_{p_1}^{\vec{n},\epsilon}$ or $\pi_{\tilde{d}_{\vec{n},\epsilon,\tilde{\epsilon}}}$ to these Hamiltonians, as we had anticipated in Sec. \ref{MS}. In the only instances in which these momenta appear, they contribute quadratically.
	
Finally, let us comment that, when the quadratic contributions to the zero mode of the Hamiltonian are explicitly computed, additional terms appear which can be reabsorbed through redefinitions of the Lagrange multipliers (see Refs. \cite{hybridGIper,tensormodes} for details). This process leads finally to a total Hamiltonian that, with an obvious notation for the modified Lagrange multipliers, can be written in the form
\begin{equation}
H=\bar{N}_0\,e^{-3\tilde{\alpha}}\tilde{H}+\sum_{\vec{n},\epsilon}G_{\vec{n},\epsilon}V_{p_2}^{\vec{n},\epsilon}+\sum_{\vec{n},\epsilon}K_{\vec{n},\epsilon}V_{p_3}^{\vec{n},\epsilon}.
\end{equation}

\section{Hybrid loop quantization}
\label{quantum}
	
In this section, we address the quantization of our model following the approach of hybrid loop quantum cosmology or, for short, hLQC \cite{hybridGIper,hybridperturbations1,hybridperturbations2,hybridperturbations3}. This quantization strategy is based on the assumption that there exists a certain regime of the quantum dynamics where the most relevant quantum geometric effects are those encoded in the zero modes of the (homogeneous) geometry, while the perturbations may be described using a more standard quantum representation. In view of this hypothesis, it seems reasonable to adopt two different quantum representations: one, of a quantum gravitational nature, for the homogeneous sector and another more conventional for the inhomogeneities, e.g. a QFT-like Fock representation. Thus, our objective is to quantize the symplectic manifold that describes our cosmological system \emph{as a whole}, employing quantum representations of different nature for the homogeneous and inhomogeneous sectors, and imposing the constraints quantum mechanically (according to Dirac's program \cite{Dirac}). 
	
We assume a quantization of the homogeneous variables that provides a representation of the canonical commutation relations such that the operators that describe the background FLRW geometry commute with the homogeneous scalar field operators (as it already happens at the level of the Poisson brackets algebra). Let these geometric and scalar field operators be defined on the kinematical Hilbert spaces $\mathcal{H}_{\rm kin}^{\rm grav}$ and $\mathcal{H}_{\rm kin}^{\rm matt}$, respectively. Then, provided our assumption on the commutation properties of the homogeneous operators, we can write the kinematical Hilbert space associated with the full homogeneous sector as the tensor product of the two mentioned representation spaces. Concerning the perturbations, we assume a Fock quantization (although this formalism is easily extensible to account for different choices \cite{hybridGIper}) such that the operators representing the basic variables for the inhomogeneities also commute with the homogeneous ones. These, together with a suitable prescription for a symmetric factor ordering upon quantization, are essentially the only building blocks necessary for the general quantum theory that was introduced in Ref. \cite{hybridGIper}, that in fact does not require at all that the methodology be particularized to a concrete quantization of the homogeneous geometry. In this sense, although in the present work we focus our attention on the case where we select a polymeric representation of the geometry, inspired by LQG, we emphasize that this is only a particular case. Whereas it is especially interesting owing to the physics emerging from LQC and the reasons discussed in the Introduction, it is by no means necessary in order to construct a formalism of hybrid quantum cosmology, that in a more general context might rest on a different quantum representation of the background degrees of freedom.

\subsection{Quantum representation of the homogeneous sector}
	
Let us now detail the quantum representation that we are going to choose for the zero mode of the scalar field and for the homogeneous FLRW geometry. We also specify the quantum counterpart of the homogeneous contribution to the zero mode of the Hamiltonian constraint, and discuss some ambiguities that appear in \emph{its definition}.
	
As far as the homogeneous matter scalar field is concerned, we consider a standard Schr\"odinger representation, in which the operator $\hat{\tilde{\varphi}}$ acts by multiplication and the momentum operator $\hat{\pi}_{\tilde{\varphi}}$ acts as a generalized derivative. The kinematical Hilbert space corresponding to the matter content is $\mathcal{H}_{\rm kin}^{\rm matt}=L^2(\mathbb{R},d\tilde{\varphi})$. 
	
Regarding the homogeneous geometry, we henceforth particularize our discussion to the case where it is described quantum mechanically by employing the formalism of LQC \cite{AS,LQC}. More concretely, we follow the so-called "improved dynamics prescription" introduced in Ref. \cite{APS2}, which accounts for the existence of a minimum nonvanishing eigenvalue $\Delta$ allowed for the area in LQG \cite{ALQG,Thiem}. Among the possible factor ordering prescriptions for the quantum representation of the Hamiltonian constraint, we adopt the symmetric prescription put forward in Ref. \cite{MMO}, usually referred to as Mart\'in-Benito--Mena Marug\'an--Olmedo (or MMO, for short) prescription. Furthermore, we want to study the application of the hybrid formalism to the particular case where the background is described using the DL procedure for the regularization of the homogeneous Hamiltonian constraint, as we will explain below.
	
In homogeneous and isotropic LQC, instead of describing the geometry using the scale factor and its conjugate momentum, the gravitational degrees of freedom are encoded in the Ashtekar-Barbero $\mathfrak{su}(2)$ gauge connection and the densitized triad, which compose a canonical pair (in the sense that their Poisson bracket is proportional to the identity). However, given the homogeneous and isotropic nature of the spatial sections of the cosmologies under study, all the relevant information is actually contained in two dynamical variables, $c$ and $p$, coming from the connection and the triad, respectively. The consideration of the improved dynamics scheme motivates a change of variables $(c,p)\rightarrow (b,v)$, where $b$  is classically proportional to the expansion rate and $v$ is the physical volume of the Universe, up to a constant multiplicative factor. This new set of variables remains canonical and their Poisson bracket is $\{b,v\}= 2$. Their precise relation to the scale factor and its canonically conjugate momentum is 
\begin{eqnarray}
e^{\tilde{\alpha}}&=&\left(\dfrac{3\gamma\sqrt{\Delta}}{2\sigma}|v|\right)^{1/3}=\left(\dfrac{3l_0}{4\pi G}\right)^{1/2}V^{1/3},\\
\pi_{\tilde{\alpha}}&=&-\dfrac{3}{2}bv,
\end{eqnarray}
where $\gamma$ is the Immirzi parameter and $V=2\pi G\gamma\sqrt{\Delta}|v|$ is the physical volume of the Universe.
	
Since the connection does not have a well-defined quantum analog in LQG, one chooses the holonomies of the connection instead as basic variables  which, together with the triad, play the role of fundamental operators in the quantum theory. The holonomy elements are given by complex exponentials of $b$. With the typical notation of the improved dynamics formulation, we denote these exponentials by $\mathcal{N}_{\pm n\bar{\mu}}=\text{exp}(\pm inb/2)$. Although we may allow $n$ to be real, in the rest of our discussion we mainly restrict our attention to integer values of $n$. The kinematical Hilbert space $\mathcal{H}_{\rm kin}^{\rm grav}$ for the geometry of flat FLRW in LQC is formed by the linear span of all the eigenstates of the volume variable,  $\ket{v}$ with $v\in\mathbb{R}$, Cauchy-completed with respect to the norm defined by the discrete inner product $\langle v|v'\rangle=\delta_{v,v'}$ \cite{LQC}. On the basis states $\ket{v}$, the volume simply acts by multiplication, whereas the holonomy operators produce constant shifts in their label, $\hat{\mathcal{N}}_{\pm n\bar{\mu}}\ket{v}=\ket{v\pm n}$. 
	
Let us now discuss the quantization of the Hamiltonian constraint of flat FLRW cosmologies in LQC, that will determine precisely the homogeneous contribution to the zero mode of the quantum Hamiltonian constraint in our perturbed model. The first step in this process is to reexpress the Hamiltonian $H_{|0}$ in terms of the basic variables of the theory, which have a well-defined quantum analog. In the LQC literature, this procedure is usually understood as a \emph{regularization} process, on account of the fact that it involves the replacement of the connection and its associated curvature tensor by holonomies around closed circuits with a nonvanishing physical area, thereby dealing with ultraviolet divergences of the classical theory. The key point in this regard is that there is not full consensus in the regularization of the Hamiltonian; indeed, different proposals exist in the literature that lead to different loop quantum theories. Actually, there are two prominent regularization proposals in LQC: the most frequent or standard one \cite{APS2} and the DL proposal \cite{DL1}. 

To comprehend the main difference between these regularization proposals, it is important to understand the basic structure of the Hamiltonian constraint in full general relativity. When written in terms of Ashtekar-Barbero variables, the Hamiltonian constraint is essentially composed by two pieces, namely the Euclidean and the Lorentzian parts, that can respectively be expressed in terms of the curvature of the connection and of the extrinsic curvature (apart from the triad), and that receive their names from the fact that only the first of these parts appears in Euclidean gravity. Traditionally, the community of LQC has employed a regularization scheme that exploits the high symmetry of the most commonly considered cosmological spacetimes, namely homogeneity and spatial flatness \cite{APS2,MMO}. Indeed, in homogeneous and spatially flat scenarios, the Lorentzian part of the Hamiltonian constraint turns out to be classically proportional to the Euclidean part and, thus, the full Hamiltonian can be expressed in terms of the Euclidean part alone. Hence, regularizing the Euclidean part suffices to rewrite the full Hamiltonian as a function of holonomies (and triads) in this kind of systems. However, conceptually, this prescription may not seem totally satisfactory, since it cannot be applied to more general scenarios, where the aforementioned symmetries fail to exist. 
	
As we have commented, recently Dapor and Liegener put forward an alternative regularization scheme which does not rely on these symmetry considerations \cite{DL1,DL2,DL3}. Indeed, it is based on the independent regularization of the two terms in the Hamiltonian constraint (through the use of two Thiemann identities). The subsequently modified model of LQC, sometimes referred to as the DL formalism of LQC, appears to lead to physical predictions which differ from the ones attained with the standard formalism of LQC, e.g. concerning the bounce mechanism that is expected to resolve the big bang singularity in the quantum theory. Results of this type raise the question of whether the standard approach to LQC faithfully captures the actual cosmological dynamics and singularity resolution picture within full LQG. In this sense, it seems enlightening to examine alternative loop quantizations of cosmological spacetimes, like the one that results in the DL formalism, in order to analyze whether the standard physical predictions are robust independently of the regularization process adopted to construct the formulation of LQC. 
	
With this motivation in mind, we now study the hybrid quantization of perturbative inhomogeneities that propagate on a homogeneous and isotropic background described by the DL formalism of LQC. The Hamiltonian constraint operator for a flat FLRW cosmology was first constructed and analyzed employing the MMO quantization prescription in Ref. \cite{DaPorLie}. As shown in that reference, the densitized version of the Hamiltonian operator for the unperturbed flat FLRW background is given by $(\hat{\pi}_{\tilde{\varphi}}^2-\hat{\mathcal{H}}_0^{(2)})/2$, with \cite{DaPorLie}
\begin{equation}\label{H02hat}
\hat{\mathcal{H}}_{0}^{(2)}=-\left(\dfrac{3}{4\pi G}\right)^2\left(\hat{\Omega}_{2\bar{\mu}}^2-\dfrac{1+\gamma^2}{4\gamma^2}\hat{\Omega}_{4\bar{\mu}}^2+\dfrac{3l_0^3}{2\pi G}\hat{V}^2\hat{\bar{W}}\right).
\end{equation}
In the previous expression, the field potential operator is to be understood as the multiplicative operator $\hat{\bar{W}}=\bar{W}(\hat{\tilde{\varphi}})$ (like any function of the zero mode of the scalar field, for that matter). The operator $\hat{\Omega}_{n\bar{\mu}}$ (for any integer number $n$) is defined as\footnote{We notice a slight modification in the quantum representation of the powers of the volume with respect to the one presented in Ref. \cite{DaPorLie}. This is due to the choice of a different prescription for the representation of the inverse of the minimum coordinate length $\bar{\mu}$. For further details, consult the Appendix of Ref. \cite{BianchiDL}.} 
\begin{equation}\label{Omega}
\hat{\Omega}_{n\bar{\mu}}=\dfrac{1}{4i\sqrt{\Delta}}\hat{V}^{1/2}\left[\widehat{\text{sgn}(v)},\hat{\mathcal{N}}_{n\bar{\mu}}-\hat{\mathcal{N}}_{-n\bar{\mu}}\right]_{+}\hat{V}^{1/2},
\end{equation}
where $[\cdot,\cdot]_{+}$ denotes the anticommutator. 
	
This operator, which is densely defined on the tensor product $\mathcal{H}_{\rm kin}^{\rm grav}\otimes\mathcal{H}_{\rm kin}^{\rm matt}$, satisfies a number of properties which are relevant to our present discussion. In the first place, it annihilates the quantum state of vanishing volume (i.e. the quantum analog of the classical singularity) and leaves invariant its orthogonal complement $\tilde{\mathcal{H}}_{\rm kin}^{\rm grav}\otimes\mathcal{H}_{\rm kin}^{\rm matt}$, where $\tilde{\mathcal{H}}_{\rm kin}^{\rm grav}$ is the Cauchy completion of the span of the volume eigenstates with a nonzero volume. This decoupling of the singular state, together with the fact that positive and negative volumes are not connected by the repeated action of the constraint, leads to the Hilbert subspaces spanned by the eigenstates with positive or negative volumes being left invariant. Furthermore, the action of the constraint superselects for the FLRW geometry Hilbert subspaces $\mathcal{H}_{\varepsilon}^{\pm}$ with support on discrete semilattices of step four $\{\pm(\varepsilon+4n), n\in\mathbb{N}\}$ \cite{DaPorLie,BianchiDL}, that have a strictly positive minimum $\varepsilon$ or a strictly negative maximum $-\varepsilon$ for the volume. For the sake of definiteness, from now on we restrict our discussion to $\mathcal{H}_{\varepsilon}^{+}$ with a fixed $\varepsilon\in(0,4]$. 

Since we  regard the inhomogeneities of our system as perturbations, it seems reasonable to demand that their introduction does not alter the superselection sectors of the unperturbed model. For this reason, we will take into account the details about the invariant Hilbert subspaces in the quantization of the quadratic contributions to the zero mode of the Hamiltonian constraint, which must preserve these subspaces as well.

\subsection{Fock representation of the inhomogeneities and implementation of the perturbative constraints}
	
For the perturbations, we consider a Fock quantization, which is selected up to unitary equivalence by the criteria of invariance of the vacuum under the spatial isometries and unitary implementability of the quantum dynamics. Additionally, we may require that the chosen Fock quantization satisfy other conditions that would further restrict the representation (for instance, we could demand that the operators constructed out of elements of the Weyl algebra and other relevant operators be well defined at the quantum level). 
	
The representation where the associated creation and annihilationlike variables correspond just to harmonic oscillators of constant frequency $\omega_n$ belongs to the family of unitarily equivalent Fock quantizations picked out by our criteria of symmetry invariance and dynamical unitarity. Let us consider this representation for simplicity, although one may instead use another representation in its equivalence class with better physical properties, according to our comments above. The corresponding Fock spaces for the MS and tensor modes are denoted by $\mathcal{F}_{S}$ and $\mathcal{F}_T$, respectively. An orthonormal basis of these spaces is given by the occupancy-number states, labeled by a positive integer per mode. Creation and annihilation operators act on this basis in the standard way, namely, by increasing or decreasing the occupancy number corresponding to a particular mode by one unit.
	
In order to complete the quantum description of the system, we still have to represent the constraints and impose them quantum mechanically. Notice that, in the gauge invariant formulation presented in the previous section, the constraints Poisson commute, a fact which allows us to impose them without the introduction of inconsistencies (at least if their quantum analogs commute as well \cite{Dirac}). Let us deal, in the first place, with the linear perturbative constraints. Our classical formalism was already designed to facilitate their imposition at the quantum level. For the part of the inhomogeneous sector parametrized by $\{V_l^{\vec{n},\epsilon}\}_{l=2,3}$, we select a quantization such that the momenta (i.e., the linear perturbative constraints under discussion) act as generalized derivatives with respect to the configuration variables $\{V_{q_l}^{\vec{n},\epsilon}\}_{l=2,3}$. In such a quantization, the vanishing of the classical constraints has a straightforward quantum counterpart: the physical states cannot depend on the configuration variables of this part of the inhomogeneous sector, since generalized derivatives with respect to them must be equal to zero. Therefore, imposing these quantum constraints amounts to the restriction to a representation space which is simply $\mathcal{H}_{\rm kin}^{\rm grav}\otimes \mathcal{H}_{\rm kin}^{\rm matt}\otimes \mathcal{F}_S\otimes\mathcal{F}_T$. Notice, however, that this Hilbert space is not the physical one yet, since the zero mode of the Hamiltonian constraint still remains to be represented and imposed (something considerably more complicated than imposing the linear perturbative constraints).
	
We focus our attention here on the \emph{densitized} version of the zero mode of the Hamiltonian constraint, $\tilde{H}$ \cite{MMO,DaPorLie}. By virtue of Eq. \eqref{Htilde} and the definition of $\hat{\mathcal{H}}_0^{(2)}$ \eqref{H02hat}, we obtain that a straightforward quantization leads to
\begin{equation}\label{Htildehat}
\hat{\tilde{H}}=\dfrac{1}{2}\left(\hat{\pi}_{\tilde{\varphi}}^2-\hat{\mathcal{H}}_0^{(2)}-\hat{\Theta}_e^{S}-\hat{\Theta}^T-\widehat{\Theta_o^S\pi_{\tilde{\varphi}}}\right),
\end{equation}
where the different operators involved will be constructed in detail in the following subsections.

\subsection{Factor ordering prescriptions}\label{rules}
	
Notice that the presence of classically noncommuting quantities in Eq. \eqref{Htildehat} (in particular, in the last three terms) makes it necessary to specify a proposal for the factor ordering that must be taken upon quantization. For the sake of clarity, we now explain and discuss the details of this proposal. We adopt the following prescriptions: 

\begin{enumerate}[label=i.]
\item The products of the form $f(\tilde{\varphi})\pi_{\tilde{\varphi}}$ are represented quantum mechanically by $\frac{1}{2}[f(\tilde{\varphi}),\hat{\pi}_{\hat{\tilde{\varphi}}}]_{+}$, where $f$ is an arbitrary function. In particular, this implies $\widehat{\Theta_o^S\pi_{\tilde{\varphi}}}=\frac{1}{2}[\hat{\Theta}_o^S,\hat{\pi}_{\tilde{\varphi}}]_{+}$ in Eq. \eqref{Htildehat}.
\end{enumerate}
	
Moreover, in the products of any real power of the volume with any function of $bv$ (that typically arises in the regularization with holonomy elements), we adopt an algebraic symmetrization for the powers of the volume (or the inverse volume). Explicitly:
\begin{enumerate}[label=ii.]
\item The products of the form $V^rg(bv)$, where $r$ is a real number and $g$ is an arbitrary function, are represented by $\hat{V}^{r/2}\hat{g}\hat{V}^{r/2}$. 
\end{enumerate}
	
The only remaining issue to be addressed is the quantization of the functions of $bv$ themselves, which requires further comments. To begin with, the quantity $bv$ is classically proportional to the  momentum variable associated with the logarithmic scale factor. The formalism presented in the previous sections provides us with a straightforward and natural definition of the quantum analog of the \emph{square} of $\pi_{\tilde{\alpha}}$. Indeed, given the definition of $\mathcal{H}_0^{(2)}$ and the DL proposal \eqref{H02hat}, we can simply set
\begin{equation}\label{hatpi2}
\hat{\pi}_{\tilde{\alpha}}^2=\hat{\mathcal{H}}_0^{(2)}+2\widehat{e^{6\tilde{\alpha}}}\hat{\bar{W}}=\hat{\mathcal{H}}_0^{(2)}+ 2\left(\dfrac{3 l_0}{4\pi G}\right)^3\hat{V}^2\hat{\bar{W}}=-\left(\dfrac{3}{4\pi G}\right)^2\left(\hat{\Omega}_{2\bar{\mu}}^2-\dfrac{1+\gamma^2}{4\gamma^2}\hat{\Omega}_{4\bar{\mu}}^2\right).
\end{equation}
Since $\hat{\Omega}_{2\bar{\mu}}$ and $\hat{\Omega}_{4\bar{\mu}}$ produce shifts of two and four units in the label of the volume eigenstates [see Eq. \eqref{Omega}], respectively, it is immediate to conclude that this quantum representation of $\pi_{\tilde{\alpha}}^2$ leaves invariant Hilbert subspaces with support on discrete lattices of step four (indeed, it is essentially the Hamiltonian constraint that would correspond to vacuum flat FLRW cosmology). 
	
We note that, while we might adopt for $\hat{\pi}_{\tilde{\alpha}}^2$ the same prescription as in Ref. \cite{hybridGIper} (that is, we might represent $\pi_{\tilde{\alpha}}^2$ as being proportional to $\hat{\Omega}_{2\bar{\mu}}^2$), that alternative definition seems less natural and convenient than the one that we have proposed above. On the one hand, even though $\hat{\Omega}_{2\bar{\mu}}^2$ is indeed a representation of the classical quantity $(bv)^2$ (up to some constant multiplicative factor), it would not agree with our choice of regularization procedure for the homogeneous Hamiltonian. Furthermore, even if we ignored this issue and admitted the use of different regularization procedures to provide quantum representations of the same object, the definition that we have proposed behaves better upon inversion. Indeed, while zero is known to belong to the spectrum of $\hat{\Omega}_{2\bar{\mu}}$ (which might lead to problems when computing the inverse), our present proposal fares better in this regard. Actually, on physical solutions to the homogeneous Hamiltonian constraint, $\hat{\mathcal{H}}_0^{(2)}$ is nonnegative and, {\sl a fortiori}, $\hat{\mathcal{H}}_0^{(2)}+2\widehat{e^{6\tilde{\alpha}}}\hat{\bar{W}}$ is also nonnegative if so is the scalar field potential, as it is often the situation in the most interesting physical scenarios (e.g., a mass term). Hence, it is not difficult to conclude that the only case in which we might encounter then a problem for physical states is at the zeros of the field potential, and only if the kinetic energy of the scalar field vanishes there quantum mechanically at zeroth-order in the perturbations, a possibility which is in any case much more stringent than the situation that we had found with the alternative representation (another argument supporting our proposed choice of representation can be found at the end of Sec. \ref{quadraticterms}).
	
Taking into account the above arguments, we choose to represent $\pi_{\tilde{\alpha}}^2$ as in Eq. \eqref{hatpi2} and, thus, any even power of $\pi_{\tilde{\alpha}}$ can be quantized in a simple way as follows:
\begin{enumerate}[label=iii.]
\item The even powers of the canonical momentum associated with the logarithmic scale factor, $\pi_{\tilde{\alpha}}^{2k}$ (for any integer $k$), are represented by 
\begin{equation}
(\hat{\pi}_{\tilde{\alpha}}^2)^k=\left(\hat{\mathcal{H}}_0^{(2)}+2\widehat{e^{6\tilde{\alpha}}}\hat{\bar{W}}\right)^k=  \left(\dfrac{3}{4\pi G}\right)^{2k}\left(-\hat{\Omega}_{2\bar{\mu}}^2+\dfrac{1+\gamma^2}{4\gamma^2}\hat{\Omega}_{4\bar{\mu}}^2\right)^k.
\end{equation}
\end{enumerate}
	
The quantum representation of the odd powers of $\pi_{\tilde{\alpha}}$, however, is not so immediate. The extra difficulty arises owing to the fact that, unlike the case of $\hat{\pi}_{\tilde{\alpha}}^2$, there is no definition of $\hat{\pi}_{\tilde{\alpha}}$ that is straightforwardly provided by the formalism. Thus, we need to introduce one ourselves. Let us denote by $\hat{\Lambda}$ the operator that results from the quantization of $\pi_{\tilde{\alpha}}$. It is obvious that any odd power of the momentum can be rewritten as an even power (for which we already have a representation) times $\pi_{\tilde{\alpha}}$ itself. Then, we are in a position to adopt an algebraic symmetrization for the even powers and represent the single remaining factor by $\hat{\Lambda}$, the form of which is yet to be discussed. As a result,
\begin{equation}
\widehat{\pi_{\tilde{\alpha}}^{2k+1}}=|\hat{\pi}_{\tilde{\alpha}}^2|^{k/2}\hat{\Lambda}|\hat{\pi}_{\tilde{\alpha}}^2|^{k/2},
\end{equation}
where $|\hat{A}|$ is the absolute value of the operator $\hat{A}$. We recall that $\hat{\pi}_{\tilde{\alpha}}^2$ is a nonnegative operator on physical states when the perturbations are absent or ignored, so that in this situation the absolute values in our definition would be spurious.
	
To conclude, let us discuss a possible way to define $\hat{\Lambda}$. A formal restriction on $\hat{\Lambda}$ is that it must only produce shifts in the volume which are integer multiples of four. Otherwise, the resulting operator would not leave invariant the superselection sectors $\mathcal{H}_{\varepsilon}^{\pm}$. In Ref. \cite{hybridGIper}, it was suggested to represent $\hat{\Lambda}$ by $\hat{\Omega}_{4\bar{\mu}}$, up to multiplicative constants, an operator which did not appear in principle in the homogenous Hamiltonian (recall that $H_{|0}$ was constructed in that work using exclusively the standard regularization), although it is often employed in LQC to represent the Hubble parameter. Indeed, this operator has the good property of respecting the superselection sectors of the unperturbed model. Since there is no longer any need to define $\hat{\Omega}_{4\bar{\mu}}$ \emph{ad hoc}, it seems reasonable to adopt the same representation of $\hat{\Lambda}$ within our DL formulation. In total, thus, we adopt the following prescription:
\begin{enumerate}[label=iv.]
\item The odd powers of the canonical momentum associated with the logarithmic scale factor $\pi_{\tilde{\alpha}}^{2k+1}$ (for any integer $k$) are represented by 
\begin{equation}
\widehat{\pi_{\tilde{\alpha}}^{2k+1}}=|\hat{\pi}_{\tilde{\alpha}}^2|^{k/2}\hat{\Lambda}|\hat{\pi}_{\tilde{\alpha}}^2|^{k/2}=\left(\dfrac{3}{4\pi G}\right)^{2k}\bigg|\hat{\Omega}_{2\bar{\mu}}^2-\dfrac{1+\gamma^2}{4\gamma^2}\hat{\Omega}_{4\bar{\mu}}^2\bigg|^{k/2}\hat{\Lambda}\ \bigg|\hat{\Omega}_{2\bar{\mu}}^2-\dfrac{1+\gamma^2}{4\gamma^2}\hat{\Omega}_{4\bar{\mu}}^2\bigg|^{k/2},
\end{equation}
where $\hat{\Lambda}$ is an appropriately chosen operator that leaves invariant the Hilbert subspaces with support on discrete semilattices of step four, such as $-3\,\hat{\Omega}_{4\bar{\mu}}/(8\pi G \gamma)$.
\end{enumerate}
	
This completes the characterization of the factor ordering prescriptions required to quantize the contributions arising from the perturbations in the zero mode of the Hamiltonian constraint. 

\subsection{Quantization of the perturbative contributions to the Hamiltonian constraint}\label{quadraticterms}
	
Let us finally analyze the result of quantizing the quadratic perturbative contributions to the Hamiltonian constraint of the system employing the proposal that we have detailed in the previous subsection. With this aim, we now represent the functions of the homogeneous phase space given in Eqs. \eqref{o}-\eqref{Tq}, as densely defined operators on $\mathcal{H}_{\varepsilon}^{\pm}\otimes L^2(\mathbb{R},d\tilde{\varphi})$.
	
We consider first the function $\vartheta_o$, defined in Eq. \eqref{o}. This is the only instance where an odd power of $\hat{\pi}_{\tilde{\alpha}}$ appears. Representing the four powers of the scale factor by the appropriate powers of the volume operator and following our prescriptions, we arrive at 
\begin{equation}
\hat{\vartheta}_o=-12l_0^2\hat{\bar{W}}'\hat{V}^{2/3}\bigg|\hat{\Omega}_{2\bar{\mu}}^2-\dfrac{1+\gamma^2}{4\gamma^2}\hat{\Omega}_{4\bar{\mu}}^2\bigg|^{-1/2}\hat{\Lambda}\ \bigg|\hat{\Omega}_{2\bar{\mu}}^2-\dfrac{1+\gamma^2}{4\gamma^2}\hat{\Omega}_{4\bar{\mu}}^2\bigg|^{-1/2}\hat{V}^{2/3},
\end{equation}
where $\hat{\Lambda}$ is the operator discussed above.
	
The remaining functions are considerably simpler. The classical function $\vartheta_e$ is nothing but the square of the scale factor and, thus, its quantum counterpart is proportional to a power of the volume operator,
\begin{equation}
\hat{\vartheta}_e=\dfrac{3l_0}{4\pi G}\hat{V}^{2/3}.
\end{equation}
	
The two remaining functions are very similar, since one is the analog of the other: $\vartheta_e^q$ is found within the context of scalar perturbations and $\vartheta_T^q$ is related to tensor perturbations instead. The scalar one, which was introduced in Eq. \eqref{eq}, is quantized as
\begin{equation}\label{hateq}
\hat{\vartheta}_e^q=\dfrac{4\pi G}{3l_0}\left[\widehat{\dfrac{1}{V}}\right]^{1/3}\hat{\mathcal{H}}_0^{(2)}\left\{19+32\pi^2G^2\left(\hat{\Omega}_{2\bar{\mu}}^2-\dfrac{1+\gamma^2}{4\gamma^2}\hat{\Omega}_{4\bar{\mu}}^2\right)^{-1}\hat{\mathcal{H}}_0^{(2)}\right\}\left[\widehat{\dfrac{1}{V}}\right]^{1/3}+\left(\dfrac{3l_0}{4\pi G}\right)^2\left(\hat{\bar{W}}''-4\hat{\bar{W}}\right)\hat{V}^{4/3},
\end{equation}
where the inverse volume operator is defined in the way which is standard in LQC, namely
\begin{equation}
\left[\widehat{\dfrac{1}{V}}\right]^{1/3}=\dfrac{3}{4\pi G\gamma\sqrt{\Delta}}\widehat{\text{sgn}(v)}\hat{V}^{1/3}\left(\hat{\mathcal{N}}_{-\bar{\mu}}\hat{V}^{1/3}\hat{\mathcal{N}}_{\bar{\mu}}-\hat{\mathcal{N}}_{\bar{\mu}}\hat{V}^{1/3}\hat{\mathcal{N}}_{-\bar{\mu}}\right).
\end{equation}
The quantum counterpart of $\vartheta_{T}^q$, defined in Eq. \eqref{Tq}, is less involved and, in fact, does not require the additional factor ordering and polymeric corrections encountered in $\hat{\vartheta}_e^q$ [see the term between the inverse volume operators in Eq. \eqref{hateq}]. Explicitly, 
\begin{equation}
\hat{\vartheta}_T^q=\dfrac{4\pi G}{3l_0}\left[\widehat{\dfrac{1}{V}}\right]^{1/3}\hat{\mathcal{H}}_0^{(2)}\left[\widehat{\dfrac{1}{V}}\right]^{1/3}-\left(\dfrac{3l_0}{2\pi G}\right)^2\hat{\bar{W}}\hat{V}^{4/3}.
\end{equation}
At this point, we can present a further argument that supports our choice of $\hat{\pi}_{\tilde{\alpha}}^2$. Classically, the first term of $\vartheta_T^q$ is obtained from the analogous term of $\vartheta_e^q$ in the limit of vanishing potential. Indeed,
\begin{equation}
19-18\dfrac{\mathcal{H}_0^{(2)}}{\pi_{\tilde{\alpha}}^2}=\dfrac{19\pi_{\tilde{\alpha}}^2-18\mathcal{H}_0^{(2)}}{\pi_{\tilde{\alpha}}^2}=\dfrac{\pi_{\tilde{\alpha}}^2+36e^{6\tilde{\alpha}}\bar{W}}{\pi_{\tilde{\alpha}}^2},
\end{equation}
that tends to one as $\bar{W}\rightarrow 0$, leading to a successful recovery of the first term of Eq. \eqref{Tq}. This relation trivially holds at the quantum level if we represent $\pi_{\tilde{\alpha}}^2$ as proposed in Eq. \eqref{hatpi2}. Nonetheless, this classical relation is violated if we keep the proposal of Ref. \cite{hybridGIper}.
	
In conclusion, we have been able to represent the functions $\vartheta_o$, $\vartheta_e$, $\vartheta_e^q$, and $\vartheta_T^q$ as densely defined operators on $\mathcal{H}_{\varepsilon}^{\pm}\otimes L^2(\mathbb{R},d\tilde{\varphi})$, which amounts to providing a quantization of the perturbative contributions to the zero mode of the Hamiltonian constraint (recall that the gauge invariant perturbative variables are represented in terms of creation and annihilation operators on $\mathcal{F}_S\otimes\mathcal{F}_T$). This, together with the quantization of the homogeneous contribution, completes the quantum representation of the whole Hamiltonian at the considered perturbative order. 
	
\section{Master constraint for the perturbations and time-dependent masses}\label{effective}

In this section, we investigate the existence of solutions to the zero mode of the Hamiltonian constraint \eqref{Htildehat} that satisfy the following ansatz of separation of variables \cite{hybridGIper}:
\begin{align} 
	\Psi=\Gamma (\tilde{\alpha},\tilde{\varphi})\psi (\mathcal{N},\tilde{\varphi}).
\end{align}	
This ansatz entails that the wave functions of the quantum states under consideration can be factorized in such a way that their dependence on the homogeneous geometry (symbolically represented by $\tilde{\alpha}$) and on the modes of the perturbations (collectively denoted by $\mathcal{N}$, that is understood to refer to the occupancy numbers of the Fock states corresponding to both the scalar and tensor perturbations) can be separated. In principle, different rates of variation might be allowed in the two factors with respect to $\tilde{\varphi}$, that in certain intervals of the evolution might be regarded as an emergent internal time; nevertheless, it is important to emphasize that these assumptions about the use of an internal clock and the relative rates of variation are not needed for the analysis and derivation of the quantum dynamics of the perturbations that we present in the following. 	

Moreover, as part of our ansatz,  we only consider states of the FLRW geometry that evolve unitarily with respect to $\tilde{\varphi}$, namely $\Gamma (\tilde{\alpha},\tilde{\varphi})=\hat{U}(\tilde{\alpha};\tilde{\varphi})\chi(\tilde{\alpha})$, where $\hat{U}$ is a unitary operator and $\chi(\tilde{\alpha})$ is to be understood as an initial condition that is normalized to the unit in $\mathcal{H}_{\rm kin}^{\rm grav}$. Introducing our ansatz in the quantum constraint equation $\hat{\tilde{H}}\Psi=0$, we arrive at the following expression:
\begin{align}\label{Htildepsi=0}
&\left\{\left((\hat{\tilde{\mathcal{H}}}_0)^2-\hat{\mathcal{H}}_0^{(2)}+[\hat{\pi}_{\tilde{\varphi}},\hat{\tilde{\mathcal{H}}}_0]\right)\Gamma\right\}\psi+2(\hat{\tilde{\mathcal{H}}}_0\Gamma)(\hat{\pi}_{\tilde{\varphi}}\psi)+\Gamma (\hat{\pi}_{\tilde{\varphi}}^2\psi)-\dfrac{1}{2}[\hat{\pi}_{\tilde{\varphi}}-\hat{\tilde{\mathcal{H}}}_0,\hat{\Theta}_o^S](\Gamma \psi)\nonumber\\&-\left\{\hat{\Theta}_e^S+\dfrac{1}{2}[\hat{\Theta}_o^S,\hat{\tilde{\mathcal{H}}}_0]_{+}+\hat{\Theta}^T\right\}(\Gamma \psi)-\hat{\Theta}_o^S\left\{\Gamma (\hat{\pi}_{\tilde{\varphi}}\psi)\right\}=0,
\end{align}
where $\hat{\tilde{\mathcal{H}}}_0$ is a self-adjoint operator defined as $\hat{\tilde{\mathcal{H}}}_0=[\hat{\pi}_{\tilde{\varphi}},\hat{U}]\hat{U}^{-1}$ .

Let us now assume that, with our choice of FLRW state, it is a good approximation to neglect quantum transitions in the homogeneous geometry mediated by the zero mode of the Hamiltonian. This implies that, when taking the inner product of the left-hand side of Eq. \eqref{Htildepsi=0} with $\Gamma$ in $\mathcal{H}_{\rm kin}^{\rm grav}$, the nondiagonal terms are irrelevant and only the expectation values on $\Gamma$ remain important. This assumption holds as long as the relative dispersions of the relevant geometric operators that appear in the zero mode of the Hamiltonian be small on the quantum state $\Gamma$ under consideration. The operators that must satisfy this condition are those accompanying the independent terms $\hat{\pi}_{\tilde{\varphi}}^2\psi$, $\hat{\pi}_{\tilde{\varphi}}\psi$, and $\psi$, in the case of this last term distinguishing also between the two existing types of contributions of the perturbative modes: either quadratic on configuration variables or on their momenta. From our previous discussion, it is immediate to check that these are just a finite number of operators, acting on the FLRW geometry. Actually, their expression in our formalism can be straightforwardly derived as in Sec. 5.3 of Ref. \cite{hybridGIper}, the only differences being the modified definitions of the $\vartheta$-operators as a result of the DL regularization and the contribution of tensor modes, which were absent in that work. 

Then, with this assumption, we obtain the master constraint equation
\begin{align}
\hat{\pi}_{\tilde{\varphi}}^2\psi+\left(2\langle\hat{\tilde{\mathcal{H}}}_0\rangle_\Gamma-\langle \hat{\Theta}_o^S\rangle_\Gamma\right)\hat{\pi}_{\tilde{\varphi}}\psi=\left[\langle \hat{\Theta}_e^S+\dfrac{1}{2}[\hat{\Theta}_o^S,\hat{\tilde{\mathcal{H}}}_0]_{+}+\hat{\Theta}^T \rangle_\Gamma+i\langle d_{\tilde{\varphi}}\hat{\tilde{\mathcal{H}}}_0-\dfrac{1}{2}d_{\tilde{\varphi}}\hat{\Theta}_o^S\rangle_\Gamma+\langle\hat{\mathcal{H}}_0^{(2)}-(\hat{\tilde{\mathcal{H}}}_0)^2\rangle_\Gamma\right]\psi,
\end{align}
that is indeed quadratic on the perturbation variables and their momenta. Here, $\langle\hat{O}\rangle_\Gamma$ denotes the expectation value on $\Gamma$ in $\mathcal{H}_{\rm kin}^{\rm grav}$ and $d_{\tilde{\varphi}}\hat{O} \equiv i[\hat{\pi}_{\tilde{\varphi}}-\hat{\tilde{\mathcal{H}}}_0,\hat{O}]$, for any operator $\hat{O}$. Notice that, with the explained procedure, all the original dependence of the zero mode of the Hamiltonian constraint on the FLRW geometry has been replaced with expectation values over this geometry, quantized according to the rules of LQC (including the definition of the kinematical Hilbert space on which the expectation values are computed).

The above equation governs the dynamics of the perturbations for states that satisfy our ansatz, accepting that the transitions in the FLRW geometry are ignorable. This equation can be reinterpreted as the result of imposing the vanishing of a constraint operator $\hat{C}_{\rm per}$ acting on $\mathcal{H}_{\rm kin}^{\rm matt}\otimes \mathcal{F}_S\otimes \mathcal{F}_T$ of the form
\begin{align}
\hat{C}_{\rm per}=\hat{\pi}_{\tilde{\varphi}}^2+D_\Gamma (\tilde{\varphi})\hat{\pi}_{\tilde{\varphi}}+E_\Gamma (\tilde{\varphi})-\langle \hat{\Theta}_e^S+\dfrac{1}{2}[\hat{\Theta}_o^S,\hat{\tilde{\mathcal{H}}}_0]_{+}-\dfrac{i}{2}d_{\tilde{\varphi}}\hat{\Theta}_o^S+\hat{\Theta}^T\rangle_\Gamma,
\end{align}
where $D_\Gamma$ and $E_\Gamma$ are two $\Gamma$-dependent functions of only the zero mode of the scalar field, the expressions of which are not relevant for our analysis. This constraint operator immediately provides us with the Heisenberg equations for the modes of the perturbations: $\hat{V}_{q_1}^{\vec{n},\epsilon}$, $\hat{V}_{p_1}^{\vec{n},\epsilon}$, $\hat{\tilde{d}}_{\vec{n},\epsilon,\tilde{\epsilon}}$, and $\hat{\pi}_{\tilde{d}_{\vec{n},\epsilon,\tilde{\epsilon}}}$. These equations are linear, and can be recast as their direct classical analogs in order to study the propagation of the gauge invariant perturbations on the FLRW background within our approximate description.

On the light of the considered densitization of the constraint and the definition of the lapse function, it is straightforward to realize that $\hat{C}_{\rm per}$ generates reparametrizations in a time $\bar{T}$ that is related classically to the coordinate time $t$ via $dt=\sigma e^{3\tilde{\alpha}}d\bar{T}$. Moreover, the form of the constraint suggests the definition of a conformal time $\eta_\Gamma$ adapted to the \emph{quantum} FLRW geometry associated with the state $\Gamma$. Indeed, one can define $l_0d\eta_\Gamma=\langle\hat{\vartheta}_e\rangle_\Gamma d\bar{T}$, which is a monotonous change of time given that $\hat{\vartheta}_e$ is a positive operator. Notice that this change of time would be meaningless if the derivative $d\eta_\Gamma/d\bar{T}$ were an operator itself. In this regard, the expectation value on $\Gamma$ plays a fundamental role in ensuring that the dynamics of the perturbations presented hereunder is mathematically well defined.

We are now in an adequate position to compute the dynamical equations for the perturbation variables by taking commutators with the generator of the evolution as discussed above, or by  taking directly Poisson brackets if we prefer to consider the classical analog of these equations. According to our comments,
\begin{align}
d_{\eta_\Gamma}V_{q_1}^{\vec{n},\epsilon}&=\dfrac{l_0}{2\langle\hat{\vartheta}_e\rangle_\Gamma}\left\{V_{q_1}^{\vec{n},\epsilon},C_{\rm per}\right\}=l_0 V_{p_1}^{\vec{n},\epsilon},\\
d_{\eta_\Gamma}\tilde{d}_{\vec{n},\epsilon,\tilde{\epsilon}}&=\dfrac{l_0}{2\langle\hat{\vartheta}_e\rangle_\Gamma}\left\{\tilde{d}_{\vec{n},\epsilon,\tilde{\epsilon}},C_{\rm per}\right\}=l_0 \pi_{\tilde{d}_{\vec{n},\epsilon,\tilde{\epsilon}}},
\end{align}
where $C_{\rm per}$ denotes the straightforward classical counterpart of $\hat{C}_{\rm per}$ (obtained by treating the scalar field momentum and the modes of the perturbations as classical quantities), and $d_{\eta_\Gamma}$ denotes the derivative with respect to the introduced conformal time \cite{hybridGIper}. Hence, calculating another derivative with respect to $\eta_\Gamma$, we obtain
\begin{align}
d_{\eta_\Gamma}^2V_{q_1}^{\vec{n},\epsilon}&=\dfrac{l_0^2}{2\langle\hat{\vartheta}_e\rangle_\Gamma}\left\{V_{p_1}^{\vec{n},\epsilon},C_{\rm per}\right\}=-\left(\tilde{\omega}_n^2+M^S\right)V_{q_1}^{\vec{n},\epsilon},\label{eqs1}\\
d_{\eta_\Gamma}^2\tilde{d}_{\vec{n},\epsilon,\tilde{\epsilon}}&=\dfrac{l_0^2}{2\langle\hat{\vartheta}_e\rangle_\Gamma}\left\{\pi_{\tilde{d}_{\vec{n},\epsilon,\tilde{\epsilon}}},C_{\rm per}\right\}=-\left(\tilde{\omega}_n^2+M^T\right)\tilde{d}_{\vec{n},\epsilon,\tilde{\epsilon}},\label{eqs2}
\end{align}
where $\tilde{\omega}_n^2=l_0^2\omega_n^2$ and $M^S$ and $M^T$ are the time-dependent masses that govern the propagation of the perturbations:
\begin{align}
M^S=l_0^2\dfrac{\langle\hat{\vartheta}_e^q\rangle_\Gamma+\frac{1}{2}\langle[\hat{\vartheta}_o,\hat{\tilde{\mathcal{H}}}_0]_+\rangle_\Gamma-\frac{i}{2}\langle d_{\tilde{\varphi}}\hat{\vartheta}_o\rangle_\Gamma}{\langle\hat{\vartheta}_e\rangle_\Gamma},\qquad M^T=l_0^2\dfrac{\langle\hat{\vartheta}_T^q\rangle_\Gamma}{\langle\hat{\vartheta}_e\rangle_\Gamma}.
\end{align}
The imaginary term in $M^S$ depends on derivatives of the scalar field potential and, in practice, it is very small in situations of physical interest \cite{hybridGIper}. Therefore, we assume that we can ignore it in the following. On the other hand, although the label $\Gamma$ has been left out for the sake of simplicity, it is clear that these masses depend on time through their dependence on the dressed FLRW geometry associated with the quantum state $\Gamma$.

The derived equations of motion are of the generalized harmonic oscillator type, with no friction and with time-dependent masses that encode the main corrections of quantum geometric nature. Furthermore, they are hyperbolic in the ultraviolet limit, where the contribution of the frequency dominates.

At this point of our discussion, it may be helpful to list concisely all the assumptions that are involved in the derivation of the presented dynamical equations \eqref{eqs1} and \eqref{eqs2} for the perturbations. To begin with, we have considered a particular ansatz for solutions to the Hamiltonian constraint, such that their wave functions can be factorized separating their dependence on the homogeneous geometry and on the perturbations. Second, on these solutions, or more specifically for their FLRW part, we have neglected the quantum transitions of the homogeneous geometry mediated by the zero mode of the Hamiltonian constraint. The validity of this assumption mathematically depends on the smallness of the relative dispersions of a finite number of concrete geometric operators on the FLRW part of the quantum state under consideration. Finally, we have also implicitly assumed that the Heisenberg equations for the modes can be directly deduced in the Fock representation that is naturally associated with the chosen Mukhanov-Sasaki and tensor variables, so that the corresponding quantum field evolution of the gauge invariant perturbations has a direct classical analog generated by $C_{\rm per}$.

To conclude, let us focus our attention on FLRW states that are highly peaked on the trajectories of effective LQC. In that case, the expectation values on $\Gamma$ can be estimated by evaluating them on such trajectories. This estimation results in the so-called \emph{scalar} and \emph{tensor effective masses}, that according to our comments adopt the following expressions within effective LQC:
\begin{align}
M^S&=l_0^2\dfrac{\vartheta_e^q+\vartheta_o\tilde{\mathcal{H}}}{\vartheta_e},\\
M^T&=l_0^2\dfrac{\vartheta_T^q}{\vartheta_e}.
\end{align}
From the expression of $\vartheta_T^q$ \eqref{Tq}, it is obvious that the tensor effective mass is unaffected by the commented potential ambiguity in the regularization of the inverse powers of the momentum of the logarithmic scale factor. However, this is not the case as regards the scalar effective mass: it contains contributions of $\vartheta_o$ \eqref{o} and $\vartheta_e^q$ \eqref{eq}, both of which depend on inverse powers of $\pi_{\tilde{\alpha}}$. The fact that these need to be regularized in order to be quantized and subsequently evaluated on the trajectories of effective LQC entails that the regularization procedure adopted for their definition does leave an imprint in the mode dynamics. Thus, an analysis of the scalar effective mass enables a direct comparison between the two proposed prescriptions.
	
\section{Conclusion}\label{conclude}

We have discussed the generalization of the hybrid approach for the quantization of cosmological perturbations around a flat FLRW universe, with compact sections, and minimally coupled with a scalar field, to possible alternative regularization schemes in LQC, showing how to combine the Fock quantization of the physical degrees of freedom of the perturbations with the quantum formalism obtained with such regularizations for homogeneous and isotropic spacetimes. 

In order to construct our formulation, we have started with a truncation of the action to second order in the perturbations of the metric and the matter fields, treating the zero modes that describe the FLRW cosmologies exactly in this procedure. Following previous work by Langlois \cite{langlois} and Refs. \cite{hybridGIper,tensormodes}, we have then adopted a set of canonical variables for the perturbations formed by gauge invariants, Abelianized perturbative constraints, and suitable canonical momenta. We have shown that this set can be completed into a canonical one for our whole cosmological system, including the sector of FLRW backgrounds as part of the total phase space. The resulting formulation permits an almost straightforward imposition of the perturbative constraints, leading to the conclusion that physical states may only depend on perturbative gauge invariants and FLRW zero modes, but are still subject to one global constraint: the zero mode of the Hamiltonian constraint of the entire perturbed cosmology. This formulation is robust and valid for any quantum description of the FLRW degrees of freedom that one decides to adopt, as far as one assumes that the canonical Poisson structure that we have obtained is preserved in the passage to quantum commutators.

Supposing that one has at hand a satisfactory Fock quantization of the gauge invariant perturbations and a consistent quantum theory for the FLRW zero modes in which one of the basic operators represent the volume of the compact spatial sections (or, alternatively, their scale factor), the hybrid quantization of the studied system essentially rests on the definition of two geometric operators that are necessary to obtain the quantum representation of the subsisting Hamiltonian constraint. The first one is the operator corresponding to the square of the canonical momentum of the logarithmic scale factor. This operator is already needed to define the Hamiltonian constraint of the unperturbed FLRW cosmology. In other words, it is an operator which is fundamental to attain a quantization of homogeneous and isotropic universes. Its explicit form depends on the regularization scheme that one adopts for the gravitational Hamiltonian, but once this regularization is chosen or fixed by suitable criteria, the natural choice is to adopt the same operator representation in the perturbative contributions to the global Hamiltonian constraint. 

The second geometric operator that appears in these perturbative terms and requires a definition is a representation of the genuine canonical momentum of the logarithmic scale factor, rather than its square. This is important because the already defined square does not contain information about the sign of the momentum. Moreover, while the operator representing the square momentum preserves by construction the superselection sectors that might exist in the homogeneous and isotropic reduction, the conservation of the superselection sectors is a requirement that one must impose on the operator corresponding to the actual momentum if one wants to regard the perturbative contributions to the zero mode of the Hamiltonian constraint as genuine perturbations, not changing the basic structure of the quantum model that describes the background. In spite of the ambiguity introduced by the choice of an operator for this momentum, one can argue that the effect in most of the  physical situations of interest is not relevant. Indeed, the operator in question is needed exclusively to quantize the only term of the perturbative contributions that contains the momentum of the homogeneous scalar field (actually in a linear way). This term appears only for scalar perturbations, and not for the tensor modes. Moreover, the term contains also a factor that is the derivative of the scalar field potential [see Eq. \eqref{o}], and that in many of the situations of interest is small, such as if the scalar field is kinetically dominated, as it is usually the most appealing scenario in LQC \cite{universe}, or in inflationary regimes driven by a cosmological constant.

Once we have represented the zero mode of the Hamiltonian constraint quantum mechanically, we have explored the existence of solutions to this quantum constraint which satisfy a particularly appealing ansatz, that permits us to separate the dependence of the wave function on the perturbations from its dependence on the homogeneous FLRW geometry. This ansatz, together with the assumption that any transition in the homogeneous geometry mediated by the constraint can be neglected in the considered states, is enough to obtain a master constraint for the perturbations that involves the expectation values of the relevant geometric operators over the quantum FLRW geometry. In this way, the main quantum effects remain encoded in these expectation values and influence the quantum evolution of the perturbations. We have discussed the time parameter with respect to which this master constraint generates evolution and we have deduced the corresponding propagation equations for the gauge invariant perturbations. These equations incorporate the quantum effects of the FLRW background via expectation values on the part of the state that describes the homogeneous geometry. We have identified the time-dependent masses that determine the behavior of the perturbations, estimated them in the regime of validity of effective LQC, and noted how they are affected by the regularization ambiguities that appear in the definition of the inverse powers of the momentum of the logarithmic scale factor. Consequently, the analysis of the properties of these masses in scenarios of physical interest seems to be a good procedure to discern the genuine effects of selecting a given representation for those geometric quantities. In this regard, our results show that the scalar mass is especially interesting, because the tensor mass does not contain any contribution that depends on the commented ambiguity.

From our exposition, we see that our formulation can be adapted essentially to any reasonable proposal for the quantization of the flat FLRW cosmologies. Given the remarkable properties of the physical states in LQC, including the avoidance of the big bang singularity (that is replaced with a bounce), our interest has been focused on this quantization procedure. According to our comments, the main ambiguity in the quantum description of the primordial perturbations within this framework is the same that one encounters in homogeneous and isotropic LQC, namely the freedom in the choice of a quantum representation for the geometric part of the Hamiltonian constraint of the FLRW universes, owing to the ambiguity in the choice of a regularization scheme. In this paper we have adopted the DL proposal for this regularization. The remaining freedom that we have encountered in our quantization process can be understood as the selection of certain prescriptions in the factor ordering and the representation of the Hubble parameter (proportional to the momentum of the logarithmic scale factor). This freedom seems much less important in the selection of a quantum theory, since the choice of an operator for the Hubble parameter only affects a term that is not physically relevant in the most interesting physical situations, as we have explained above, and because one would expect that the factor ordering should not affect the fundamental properties of the formalism. 
	
\acknowledgments
	
The authors are grateful to B. Elizaga Navascu\'es for discussions. This work has been supported by Project. No. FIS2017-86497-C2-2-P of MICINN from Spain. The project that gave rise to these results received the support of a fellowship from ``la Caixa'' Foundation (ID 100010434). The fellowship code is LCF/BQ/DR19/11740028.

\end{document}